\documentclass[9pt,twocolumn,twoside]{arxiv-preprint}
\usepackage{subcaption}
\ifdefined\XeTeXversion
  \usepackage{fontspec}
\fi

\ifdefined\XeTeXversion\verticaladjustment{-12pt}\fi % XeTeX first-page compatibility.

\articletype{ASTRONOMY}%%%% Article topic/classification

\templatetype{pnasresearcharticle} % Choose template
\begin{document}

\title{The Survival Paradox of Close-In Planets Orbiting Red Clump Giants}

% Use letters for affiliations, numbers to show equal authorship (if applicable) and to indicate the corresponding author

\author[1]{Wen-Xu Lin}
\author[1*]{Sheng-Bang Qian}
\author[1]{Qi-Bin Sun}
\author[2,3]{Li-Ying Zhu}
\author[2,3]{Wen-Ping Liao}
\author[4]{Ting Li}
\author[1]{Fu-Xing Li}
\author[2,3]{Xiang-Dong Shi}
\author[1]{Tao Wang}
\author[2,3]{Ergang Zhao}
\author[1]{Huiting Zhang}
\author[1]{Ping Li}

\affil[1]{Department of Astronomy, School of Physics and Astronomy, Yunnan University, Kunming, 650091, P. R. China}

\affil[2]{Yunnan Observatories, Chinese Academy of Sciences, Kunming, 650011, P. R. China}

\affil[3]{University of Chinese Academy of Sciences, No.1 Yanqihu East Rd, Huairou District, Beijing, 101408, P. R. China}

\affil[4]{Yunnan Training Center for Ethnic Leadership, Yunnan Minzu University, Kunming, 650504, P. R. China}

% Please give the surname of the lead author for the running footer
\leadauthor{Lin \& Qian}

% Please add a significance statement to explain the relevance of your work
\significancestatement{It is commonly assumed that planets on close orbits cannot survive once their host stars expand into red giants, as stellar radii increase dramatically and tidal forces become destructive. This expectation underpins current views of planetary populations around evolved stars. While a single such system had been reported previously, we show that this phenomenon is not an isolated anomaly. The identification of multiple close-in planets orbiting red-clump giants—stars that have already experienced their maximum expansion—indicates the emergence of a previously unrecognized class of planetary systems. These systems imply the action of a common physical process operating beyond current theoretical frameworks and fundamentally reshape our understanding of stellar--planetary coevolution in the late stages of stellar evolution.}

% Please include corresponding author, author contribution and author declaration information
\authorcontributions{W.-X.L. (first author) identified the red-clump planet-host candidates, led the asteroseismic screening and dipole mixed-mode analysis, performed the radial-velocity reanalysis, derived the stellar and planetary parameters, carried out the multi-model TRGB comparison and representative orbital-evolution calculation, prepared the figures and tables, and wrote the main manuscript and revision. 
	S.-B.Q. (corresponding author) supervised the project, defined the overall scientific direction, provided resources and technical support, and contributed to the interpretation of the results.
	L.-Y.Z., W.-P.L., and F.-X.L. contributed to the radial-velocity analysis, orbital-parameter calculations, and checks of the planetary interpretation.
	Q.-B.S., X.-D.S., T.W., E.Z., H.Z. and P.L. contributed to data collection, literature comparison, and validation of the stellar and planetary parameters.
	T.L. helped refine the manuscript structure, language, and presentation. All authors discussed the results and reviewed the manuscript.}
\authordeclaration{The authors declare no competing interest.}
%\equalauthors{\textsuperscript{1}A.O.(Author One) contributed equally to this work with A.T. (Author Two) (remove if not applicable).}
\correspondingauthor{\textsuperscript{*} E-mail: qiansb@ynu.edu.cn}

% At least three keywords are required at submission. Please provide three to five keywords, separated by the pipe symbol.
\begin{abstract}
Close-in planets orbiting core-helium-burning red-clump (RC) giants directly challenge tidal-engulfment theory, which predicts that planets within about 1 au of Sun-like stars should be engulfed during red-giant-branch (RGB) expansion. The existence of such survivors would imply missing physical processes in the late stages of stellar and planetary evolution. Yet only one such “forbidden” planet had been identified to date.
Here we report three systems—24 Boo b, HD 12648 b, and TYC 4282-605-1 b—whose host stars are confirmed red-clump giants through asteroseismology, showing that each has already undergone dramatic expansion and contraction following helium ignition. Their planets now reside at orbital separations smaller than the maximum stellar radii attained during the RGB phase, implying that they should not have survived. Among them, 24 Boo b orbits at only 0.183 au with a 30-day period, making it the closest planet known around a red-clump giant.
These discoveries demonstrate that this paradox is not an isolated anomaly but signals the emergence of a new class of evolved planetary systems. Their survival defies current models of stellar–planetary coevolution and points to unexplored physics shaping planetary architectures around post-RGB stars.
\end{abstract}

\maketitle
\thispagestyle{firststyle}
\ifthenelse{\boolean{shortarticle}}{\ifthenelse{\boolean{singlecolumn}}{\abscontentformatted}{\abscontent}}{}

\Firstpage

%The \Firstpage command is used to format the first page text column size. The same size will be maintained for subsequent paragraph until the \Endparasplit or \Parasplit command is encountered.
\dropcap{C}lose-in planets orbiting evolved giant stars are generally expected to be scarce. Stellar evolution models indicate that as a star climbs the red-giant branch, its outer envelope can expand by orders of magnitude, driving tidal interactions and mass loss that are predicted to engulf any planet within a certain orbital radius \cite{villaver2009orbital, mustill2012foretellings, veras2016post}. Observationally, surveys of evolved stars show a dearth of short-period giant planets, consistent with theoretical expectations \cite{jones2014properties, reffert2015precise}. In this context, finding a planet in a close orbit around a confirmed core-helium-burning red-clump giant would be highly surprising and would challenge our understanding of how planets survive post–main-sequence evolution. 
In this study, we performed an asteroseismic screening of all known planet-hosting giant stars, through which we first identified three systems—24 Boo, HD 12648, and TYC 4282-605-1—as red-clump (RC) stars that nevertheless still harbour close-in planets. Their planets reside at orbital distances that standard interaction theory predicts would not survive the host’s prior expansion and contraction phases.

\begin{figure*}[!t]
	\centering
	\includegraphics[width=18cm]{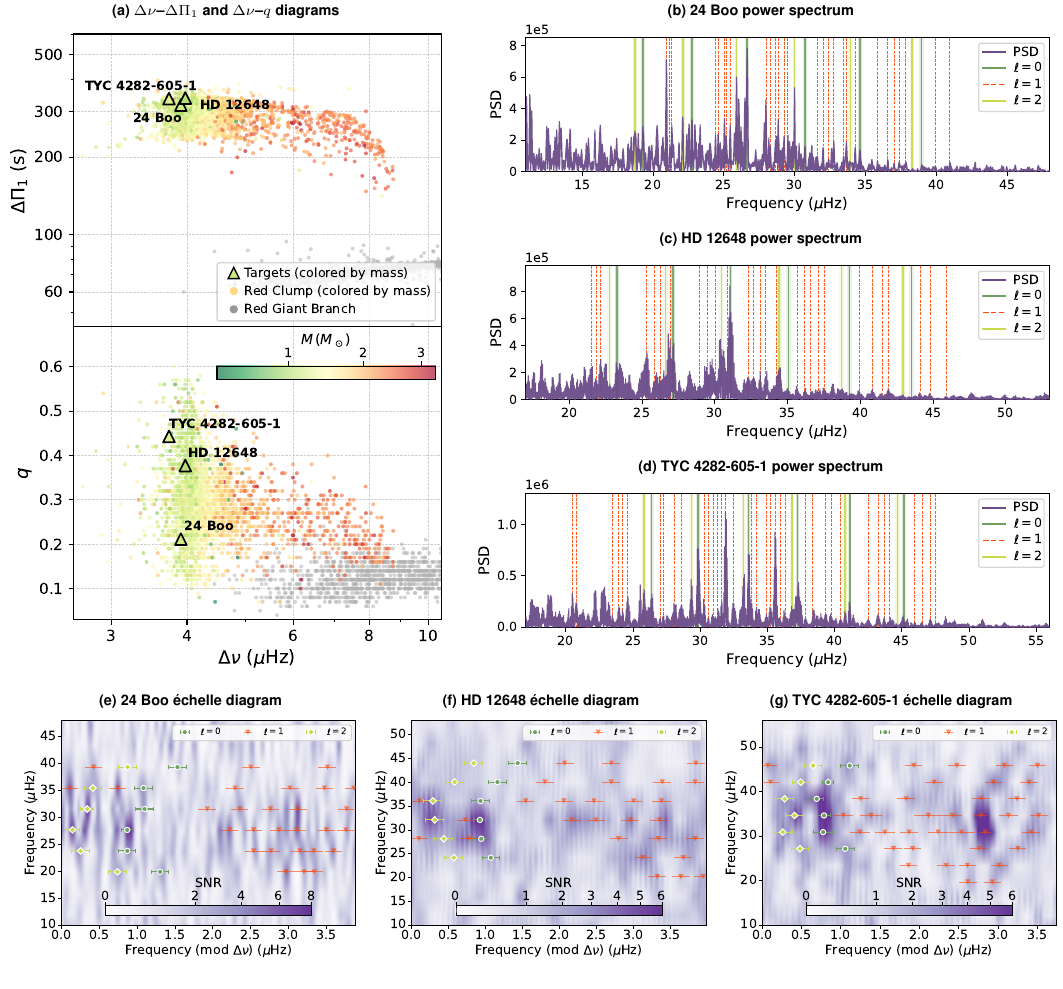}
	\caption{\textbf{Asteroseismic diagnostic diagrams for the three planet-host stars.} (a) Locations of the three host stars in the asteroseismic parameter spaces defined by the large frequency separation, $\Delta\nu$, the mode-coupling factor, $q$, and the dipole gravity-mode period spacing, $\Delta\Pi_{1}$. The host stars are marked with triangles and labeled by name, with the symbol color indicating the stellar mass. The background points show literature measurements \cite{mosser2017period} for evolved stars: red-clump stars are color-coded by mass, whereas red-giant-branch stars are shown in gray. The positions of the three host stars are clearly consistent with the red-clump population. (b--d) Observed power spectral densities with the \texttt{PBjam 2.0} radial ($\ell=0$), dipole ($\ell=1$), and quadrupole ($\ell=2$) mode pattern overlaid for 24 Boo, HD 12648, and TYC 4282-605-1, respectively. (e--g) Corresponding frequency-\'{e}chelle diagrams. The prior screening, sequential mode analysis, and separate $\ell=0,2$ and $\ell=1$ joint posterior distributions are documented in Supplementary Methods B and Supplementary Figs. S2--S8.}
	\label{fig:deltapi-dnu}
\end{figure*}

The evolutionary classification of evolved planet-hosting stars has long been uncertain, as spectroscopic and isochrone-based methods cannot reliably distinguish red-giant-branch (RGB) stars from core–helium-burning RC stars \cite{masseron2017spectroscopic, bovy2014apogee, girardi2016red, banks2024role}. This ambiguity has hindered efforts to connect the observed population of planets around giants with theoretical expectations for post–main-sequence evolution. Asteroseismology has resolved this problem by providing direct diagnostics of the stellar core through oscillation modes that probe the internal density and composition profile \cite{bedding2011gravity, mosser2014mixed, vrard2016period}. 
Applying these diagnostics, we carried out an asteroseismic screening of all currently known exoplanet host stars \cite{lin2025asteroseismic} and, for the evolved planet hosts observed by {\it TESS} \cite{jenkins2016tess}, we identified three exceptional systems in which the host stars are securely classified as core–helium–burning RC giants (Fig.~\ref{fig:deltapi-dnu}). The measurement procedure is described in Materials and Methods, with supporting light-curve, mixed-mode, and posterior diagnostics provided in the Supplementary Information. This confirmation removes the remaining ambiguity about their evolutionary stage. \Endparasplit

\noindent This classification establishes that their stars have already passed through the helium flash and the tip of the red-giant branch. Consequently, the planets in these systems are now known to orbit stars that once reached their maximum radii—an insight that transforms how we interpret their survival.

To ensure consistency across all targets, we reanalyzed the radial-velocity (RV) data for the three systems using a uniform procedure that combines all available high-precision measurements (see Materials and Methods). The orbital fits yield a coherent Keplerian solution for each close-in giant planet. Combining the RV posteriors with the asteroseismic stellar-mass posteriors gives semi-major axes of $a = 0.1827 ~\mathrm{au}$ for 24~Boo~b, $a = 0.5157 ~\mathrm{au}$ for HD~12648~b, and $a = 0.4885~\mathrm{au}$ for TYC~4282-605-1~b. Because RV measurements alone cannot constrain orbital inclination, the exact planetary masses remain uncertain; nevertheless, the inferred minimum masses place all three companions in the giant-planet or substellar-companion regime (Table~\ref{tab:para}). The remarkably small orbital separations suggest that these planets likely reside within the theoretical “forbidden zone” for post–main-sequence systems. To test this hypothesis, we constructed stellar and orbital evolution simulations to provide theoretical verification of their survival limits.

\begin{figure*}[!t]
	\centering
	\includegraphics[width=18cm]{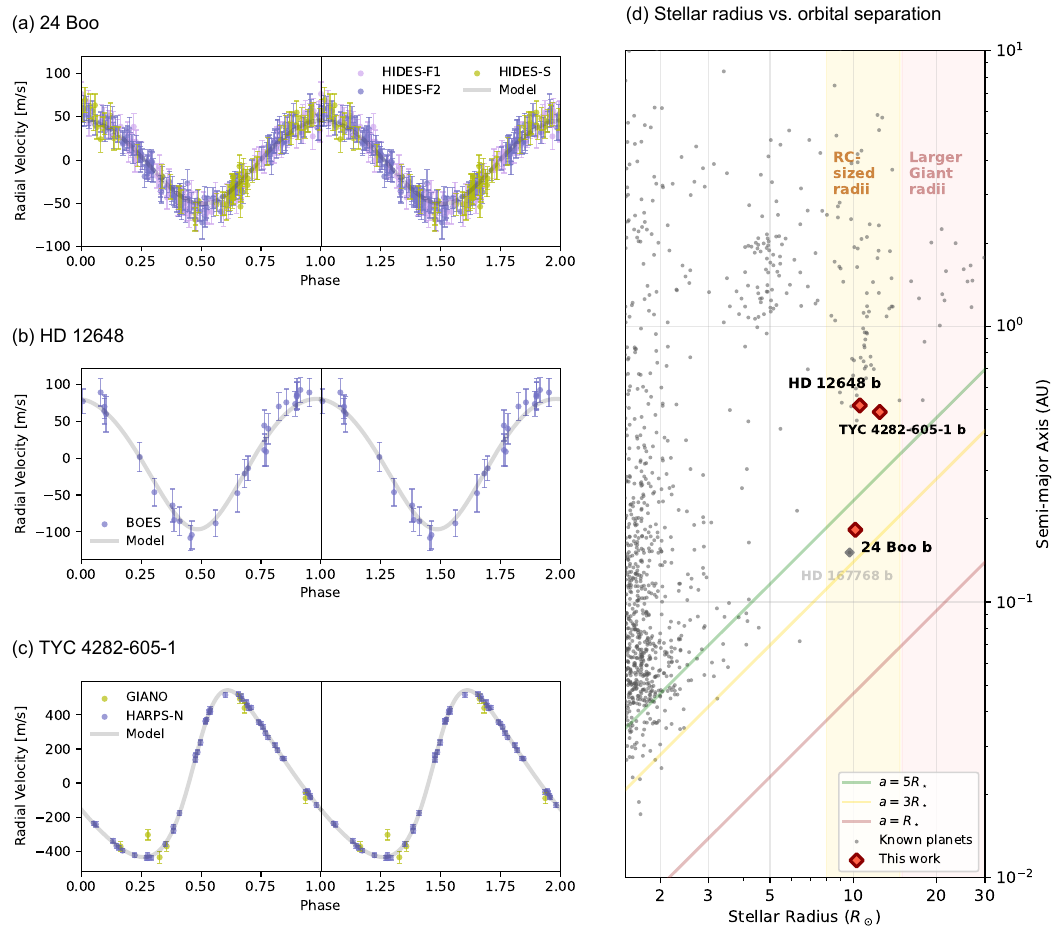}
	\caption{
		\textbf{Radial-velocity solutions and orbital locations of the three planets.}
		\textbf{(a--c)} Phase-folded radial-velocity curves for 24 Boo, HD 12648, and TYC 4282-605-1. Points show the RV measurements from different instruments, and gray curves show the best-fitting Keplerian models.
		\textbf{(d)} Stellar radius versus orbital separation for confirmed planets in the NASA Exoplanet Archive \cite{akeson2013nasa,christiansen2025nasa} whose hosts have reported radii. The yellow band marks the approximate radius range of red-clump-sized giants, and the pale red region indicates larger giant radii. Diagonal lines show $a=R_\star$, $3R_\star$, and $5R_\star$. The three planets studied here are highlighted with red diamonds. HD~167768~b \cite{2023PASJ...75..169T} is shown separately as a gray diamond because its host has an RC-sized radius but could instead be an RGB star; see Supplementary Methods H and Fig. S25.
	}
	\label{fig:DeltaPi}
\end{figure*}

\renewcommand{\arraystretch}{2.0}
\begin{table*}[!t]
	\centering
	% \footnotesize
	\small
		\caption{Parameters of Three Red-Clump Giants and Their Planetary Companions\label{tab:para}}
	\begin{tabular}{@{}lccc@{}}
		\hline\hline
		\multicolumn{4}{c}{\textbf{Parameters of host stars}}\\
		& 24 Boo & TYC 4282-605-1 & HD 12648  \\
		\hline
		$\nu_{\max}$ ($\mu\mathrm{Hz}$)   & $28.63^{+0.69}_{-0.57}$ & $33.58^{+0.44}_{-0.53}$ & $31.19^{+0.82}_{-0.71}$  \\
		$\Delta\nu$ ($\mu\mathrm{Hz}$)     & $3.908\pm0.054$  & $3.735\pm0.030$  & $3.974\pm0.038$  \\
		$\Delta\Pi_{1}$ (s)                & $318.0^{+0.4}_{-0.6}$ & $337.0^{+0.3}_{-0.4}$ & $338.8^{+0.5}_{-0.5}$\\
		$q$                                & $0.211^{+0.048}_{-0.047}$ & $0.443^{+0.082}_{-0.076}$ & $0.377^{+0.060}_{-0.061}$ \\
		$T_{\rm eff}\ (\mathrm{K})^{\dagger}$ 
		& $4893\pm122$    
		& $4505\pm122$  
		& $4835\pm122$ \\
		$[{\rm Fe/H}]^{\dagger}$ 
		& $-0.77\pm0.03$    
		& $-0.07\pm0.16$  
		& $-0.57\pm0.02$ \\
		$M_{\star}\ (\rm M_{\odot})$ 
		& $0.884_{-0.108}^{+0.118}$ 
		& $1.492_{-0.162}^{+0.168}$ 
		& $1.032_{-0.123}^{+0.133}$\\
		$R_{\star}\ (\rm R_{\odot})$ 
		& $10.173_{-0.493}^{+0.503}$ 
		& $12.471_{-0.506}^{+0.503}$ 
		& $10.561_{-0.468}^{+0.479}$\\
		$\log{g}$ ($\log_{10}\,\mathrm{cm\,s^{-2}}$) 
		& $2.369_{-0.016}^{+0.016}$ 
		& $2.420_{-0.014}^{+0.014}$ 
		& $2.404_{-0.017}^{+0.017}$\\
		$L_{\star}$ ($\log_{10}\,\rm L_{\odot}$) 
		& $1.730_{-0.086}^{+0.076}$ 
		& $1.761_{-0.078}^{+0.073}$ 
		& $1.742_{-0.080}^{+0.071}$\\
		\hline
		\multicolumn{4}{c}{\textbf{Parameters of exoplanets}}\\
		& 24 Boo b & TYC 4282-605-1 b & HD 12648 b \\
		\hline
		$P$ (days)           &  $30.337_{-0.005}^{+0.006}$    &   $101.630_{-0.057}^{+0.058}$   & $132.93_{-0.55}^{+0.71}$ \\
		$K$ (m/s)            &   $54.08_{-3.26}^{+3.10}$   &   $490.8_{-4.9}^{+4.5}$    & $99.4_{-9.5}^{+10.2}$    \\
		$t_{P}$ (BJD)       &  $2456129.8_{-2.3}^{+2.2}$  & $2456785.7_{-0.6}^{+0.7}$  & $2456365.7_{-10.5}^{+10.8}$   \\
		$e$                  &   $0.050_{-0.049}^{+0.013}$    &    $0.262_{-0.006}^{+0.007}$  & $0.091_{-0.090}^{+0.022}$ \\
		$\omega$ (deg)       &  $177.9_{-27.5}^{+26.7}$  &  $295.4_{-2.5}^{+2.6}$     & $185.7_{-26.2}^{+29.6}$   \\
		$a$ (au)             & $0.1827_{-0.0078}^{+0.0078}$ & $0.4885_{-0.0183}^{+0.0176}$  & $0.5157_{-0.0214}^{+0.0213}$   \\
		$M_{p}\sin i$ ($\rm M_{J}$) & $0.762_{-0.076}^{+0.083}$ & $14.278_{-1.051}^{+1.062}$ & $2.539_{-0.313}^{+0.343}$  \\
		\hline\hline
	\end{tabular}
	\begin{flushleft}
	\footnotesize
	\textsuperscript{$\dagger$}The effective temperatures and metallicities were adopted as external input constraints for the stellar-parameter inference. For 24 Boo, both $T_{\rm eff}$ and $[{\rm Fe/H}]$ are from Takarada et al. \cite{2018PASJ...70...59T}. For TYC 4282-605-1, $T_{\rm eff}$ is from Stassun et al. \cite{2019AJ....158..138S}, while $[{\rm Fe/H}]$ is from Gonz\'alez-\'Alvarez et al. \cite{2017AandA...606A..51G}. For HD 12648, $T_{\rm eff}$ and $[{\rm Fe/H}]$ are from Lee et al. \cite{2015A&A...584A..79L}. The adopted $122\,{\rm K}$ uncertainty on $T_{\rm eff}$ is not the formal uncertainty quoted in the original literature measurements; it is a uniform, conservative uncertainty motivated by the rms scatter of the temperature calibration in the revised TESS Input Catalog \cite{2019AJ....158..138S}.
	The quoted stellar-parameter intervals are the 16th--84th percentile ranges after propagating the observational constraints, the \texttt{asfgrid} $f_{\Delta\nu}$ correction, and the literature-informed population prior on $f_{\nu_{\max}}$. They remain conditional on this framework and do not exhaust uncertainties from unvaried stellar input physics or a possible common RC zero-point offset in $f_{\nu_{\max}}$. The semi-major axes and minimum masses are likewise reported as medians and 16th--84th percentile intervals from the Monte Carlo propagation described in Materials and Methods.
	\end{flushleft}

\end{table*}

The baseline \texttt{asfgrid} posteriors yield masses of $0.884^{+0.118}_{-0.108}$, $1.492^{+0.168}_{-0.162}$, and $1.032^{+0.133}_{-0.123}\,M_\odot$ and radii of $10.173^{+0.503}_{-0.493}$, $12.471^{+0.503}_{-0.506}$, and $10.561^{+0.479}_{-0.468}\,R_\odot$ for 24 Boo, TYC 4282-605-1, and HD 12648, respectively. The corresponding prior-width sensitivity test is reported in Supplementary Table S3. The literature comparison reveals method-dependent differences not captured by the conditional posterior intervals alone (Supplementary Table S8). For 24 Boo, the interferometric radius and luminosity from Karovicova et al. \cite{karovicova2020benchmark} agree with the seismic values inferred here, whereas the mass and derived surface gravity remain discrepant at approximately $3.1\sigma$ and $4.6\sigma$, respectively. We therefore also tested whether the low inferred RC masses are compatible with the time available for single-star evolution. A MIST grid evaluated in current RC mass and metallicity gives times from stellar birth to a representative stable RC state of 9.03, 2.55, and 6.02~Gyr for 24 Boo, TYC 4282-605-1, and HD 12648, respectively (Supplementary Fig. S17). The central constraints are thus compatible with a single-star evolutionary history, including for 24 Boo, although this consistency check does not exclude mass transfer, merger, or other evolutionary pathways.

\begin{figure*}[htbp!]
	\centering
	\includegraphics[width=18cm]{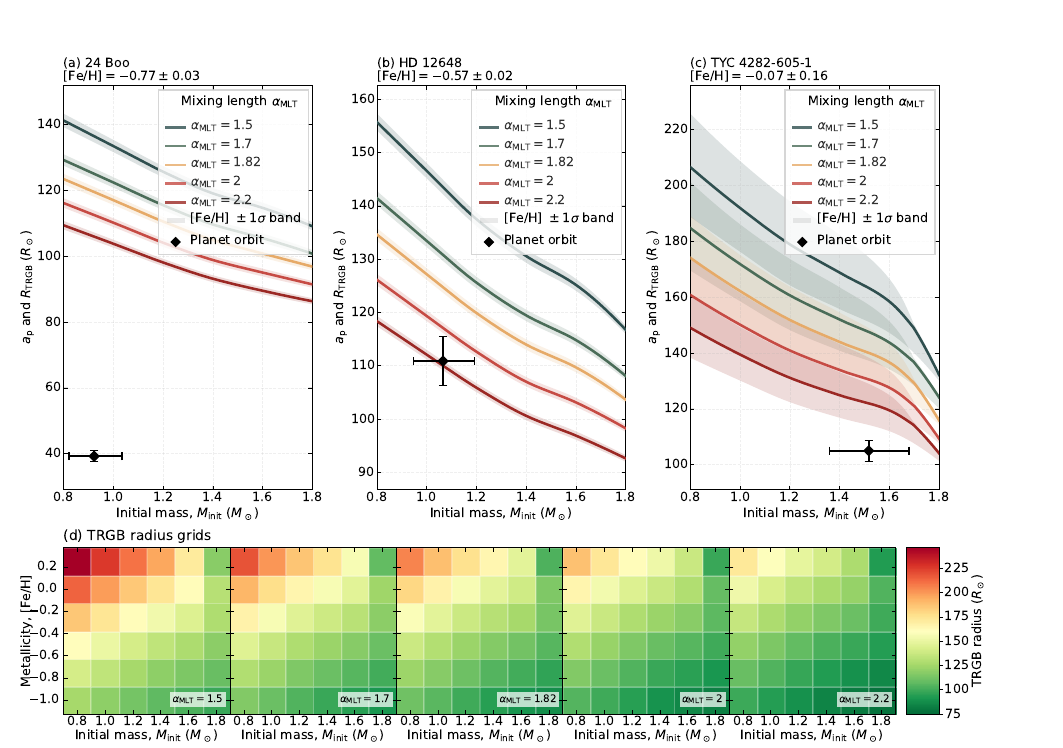}
	\caption{\textbf{Controlled sensitivity of the TRGB radius to the mixing-length parameter.} Panels (a--c) compare the measured planetary orbit (black diamond, with asymmetric orbital and mapped initial-mass uncertainties) with the MIST/MESA $R_{\rm TRGB}(M_{\rm init})$ curves calculated at the measured metallicity of 24 Boo, HD 12648, and TYC 4282-605-1. Colours denote $\alpha_{\rm MLT}=1.5$, 1.7, 1.82, 2.0, and 2.2; the shaded envelopes span each host's quoted $[\mathrm{Fe/H}]\pm1\sigma$ interval. Because the orbit and TRGB radius are plotted on the same vertical scale in solar radii, a curve above the diamond has $R_{\rm TRGB}>a_{\rm p}$ and corresponds to geometric engulfment. Panel (d) shows the underlying $6\times6$ TRGB-radius grids in initial mass and metallicity for the five mixing lengths. The solar-calibrated reference is $\alpha_{\rm MLT}=1.82$.}
	\label{fig:HR}
\end{figure*}

In standard post–main-sequence evolution theory, planets orbiting within a few astronomical units are expected to undergo orbital decay and eventual engulfment due to the combined effects of stellar expansion, tidal dissipation, and wind drag \cite{villaver2009orbital, kunitomo2011planet, mustill2012foretellings,veras2016post}. To test whether the observed systems contradict these expectations, we first used controlled MIST/MESA grids to compare each measured planetary orbit directly with $R_{\rm TRGB}(M_{\rm init})$ while varying the mixing-length parameter and the metallicity within each host's quoted uncertainty interval (Fig.~\ref{fig:HR}). Before this comparison, the present-day asteroseismic RC masses were mapped to initial masses using the MIST RC mass-loss relation documented in Supplementary Methods D and Fig. S14. The orbit of 24 Boo b lies below every tested TRGB-radius curve, and the same is true for TYC 4282-605-1 b over its measured mass range. HD 12648 b is more sensitive to the convection prescription: its grid-conditioned geometric engulfment probability decreases from nearly unity at low $\alpha_{\rm MLT}$ to 43.10\% at $\alpha_{\rm MLT}=2.2$. This controlled convection test produces a larger displacement of the survival boundary than the inter-library spread. The complementary fixed-metallicity comparison among six public stellar-evolution libraries is shown in Supplementary Fig. S15, and the controlled helium-enrichment test is shown in Supplementary Fig. S16. The full uncertainty propagation and geometric engulfment probabilities are reported in Supplementary Methods D and Table S5.

As the final step of the argument, we used a representative single-star orbital-evolution calculation to identify the limiting orbit expected for a planet that survives from the RGB into the RC phase. A binary-search procedure was used to find the critical initial orbit that just avoids engulfment at the TRGB. Thus, under canonical single-star orbital evolution, a planet observed around an RC star should not have a present-day semi-major axis smaller than the host star's TRGB radius.

The presence of close-in planets orbiting red-clump giants challenges our current understanding of stellar and planetary evolution, suggesting that additional physical mechanisms must operate beyond the canonical single-star scenario. Several evolutionary pathways may account for the survival—or reappearance—of these planets. The channels discussed below are therefore intended as qualitative possibilities rather than as mechanisms quantitatively demonstrated in this work.

\textbf{Binary or merger pathways.} If any of the host stars originated in close binary systems, past mass transfer or merger events could have modified their angular-momentum distribution and envelope expansion history, thereby stabilizing inner planetary orbits. Binary interaction has been shown to suppress envelope inflation or to delay red-giant expansion, extending the survival region for inner companions \cite{privitera2016star, stephan2020eating}. The unusually low mixed-mode coupling factor of 24 Boo contrasts with the stronger coupling generally expected at its low metallicity \cite{jiang2020variations,kuszlewicz2023mixed}, suggesting an evolutionary history that may differ from canonical single-star evolution. This possibility was explored in detail by Hon et al. \cite{hon2023close} as a possible explanation for 8 UMi b. Moreover, recent hydrodynamic simulations and observations reveal that post-merger objects can retain or reform circumstellar disks, within which “second-generation” planets might condense \cite{hirai2021simulating}. Such a scenario would naturally explain the presence of apparently close-in planets around post-flash giants whose envelopes have been reorganized by prior binary evolution.

\textbf{Post-Engulfment Planetary Disk (PEPD) hypothesis.} Multiple-planet systems are ubiquitous in the Galaxy \cite{zhu201830}. When an inner planet is engulfed during stellar expansion, its angular momentum and orbital energy can be deposited into the stellar envelope, potentially triggering envelope ejection and the formation of a transient, rotationally supported disk \cite{schleicher2014planet, hirai2021simulating, macleod2018planetary}. Related post--main-sequence disk scenarios, including second-generation planet formation in evolved binary systems, have been discussed in previous work \cite{perets2011planets,hon2023close}. In this context, a post-engulfment disk could, in principle, interact with a surviving outer planet and contribute to inward migration after the host contracts toward the red-clump phase. This scenario is related to, but distinct from, second-generation planet formation: the disk-like structure considered here would be produced by the engulfment of an inner planet, and its possible role would be to reshape the orbit of an already existing outer planet. We do not model this channel quantitatively in this work, and assessing its viability will require dedicated hydrodynamical and orbital-evolution calculations.

\textbf{Partial engulfment and envelope retreat.} A related speculative possibility is partial engulfment followed by rapid envelope contraction. In this picture, a planet may enter the extended envelope near the TRGB, but if the envelope contracts rapidly after the helium flash while the orbital-decay timescale remains longer than the envelope-contraction timescale, the planet could effectively emerge from the envelope and remain on a close orbit during the red-clump phase. Similar ideas have been discussed in the context of planets interacting with rapidly evolving envelopes near the core helium flash, where the competition between envelope evolution and inspiral timescales can allow a massive planet or substellar companion to survive a common-envelope-like interaction \cite{bear2011possible,bear2021rapid,merlov2021red}.

\textbf{Other possibilities.} 
Additional, though less probable, scenarios include the dynamical capture of free-floating planets \cite{mustill2011general}, gravitational scattering within multi-planet systems that reshuffles orbital energy after the red-giant phase \cite{rasio1996dynamical, carrera2019formation}, or perturbations from hierarchical stellar companions. While none of these explanations alone can fully reproduce the observed configuration, they highlight the diversity of potential evolutionary channels leading to “forbidden” planets around evolved giants.

The discovery of close-in planets around red-clump giants reveals a fundamental tension in our current understanding of planetary system evolution. These planets occupy orbits that should have been erased during the stars’ previous red-giant expansion, yet they persist in defiance of canonical stellar-evolution theory. By coupling asteroseismic diagnostics with detailed stellar and orbital modeling, we have shown that the host stars—24~Boo, HD~12648, and TYC~4282-605-1—have already undergone helium ignition and envelope contraction, confirming that their planets must have survived (or reformed) after the helium flash. Their existence therefore demands a re-evaluation of how tidal dissipation, mass loss, and binary interaction sculpt planetary architectures around evolved stars.  
Beyond challenging long-held theoretical boundaries, these systems provide a unique empirical bridge linking planetary formation, migration, and survival across multiple stellar evolutionary phases. They open a new observational window into the physics of angular-momentum transfer, envelope recycling, and post-engulfment disk dynamics. As the sample of such post-flash close-in planets expands, these systems will make it possible to test which of these processes contribute to planetary survival and to calibrate tidal and wind-driven evolution models.

% Figure 3

\matmethods{
\subsection*{Asteroseismic analysis}\label{sec:astero}

From the Mikulski Archive for Space Telescopes (MAST) archive, we downloaded the SPOC-pipeline \cite{jenkins2016tess} TESS light curves for three host stars (24 Boo, HD 12648 and TYC 4282-605-1). From these light curves, we computed their power spectra. The light curves and photometric periodograms used to check the RV periods are shown in Supplementary Fig. S1.
We obtained the $\nu_{\max}$ and central $\Delta\nu$ values from the \texttt{PBjam 2.0} radial--quadrupole ($\ell=0,2$) analysis; the $\Delta\nu$ uncertainty was evaluated independently from the peakbagged radial and quadrupole mode frequencies. The stellar masses, radii, surface gravities, and luminosities were then inferred with the grid-based \texttt{asfgrid} procedure \cite{sharma2016stellar,sharma2016asfgrid,stello2022extension}, using $\nu_{\max}$, $\Delta\nu$, $T_{\rm eff}$, and $[{\rm Fe/H}]$ as input constraints. In each Monte Carlo realization, these input observables and a population-based prior on the $\nu_{\max}$ scaling-relation factor $f_{\nu_{\max}}$ were sampled and passed to \texttt{asfgrid}, with the evolutionary state specified as post-RGB for the three red-clump host stars.

The model-dependent correction supplied by \texttt{asfgrid} is the correction to the large frequency separation, $\Delta\nu$, through $f_{\Delta\nu}$. To account for residual systematics in the $\nu_{\max}$ scaling relation, we also propagated an uncertainty prior on $f_{\nu_{\max}}$. Detailed modelling has shown departures from unity for evolved stars, including an approximately 10\% observed--model discrepancy for two very metal-poor first-ascent giants and broader population-level dependences on stellar properties \cite{larsen2025metalpoor,2022ApJ...927..167L}. Because neither study supplies a direct RC-specific calibration for the metallicities and parameters of our three hosts, we did not apply an extrapolated target-specific correction. Instead, we constructed a median-normalized, two-sided Gaussian prior from the empirical distribution presented by Li et al. (Supplementary Fig.~S9), with $f_{\nu_{\max}}=1.000$ and 16th and 84th percentiles of 0.9782 and 1.0218. This factor was drawn in every Monte Carlo realization, thereby incorporating the corresponding population-level systematic uncertainty into the stellar and planetary parameter posteriors. Because normalization fixes the prior median at unity, it does not capture a possible zero-point displacement common to RC stars; such an RC-specific systematic may remain. Repeating the full calculation with one, two, and three times the baseline prior width changes the stellar-parameter medians by less than 0.7\% while broadening their credible intervals as expected (Supplementary Table S3). We adopt the baseline-width posterior for the reported parameters because its width and two-sided Gaussian form are determined from the empirical distribution in Supplementary Fig.~S9; the $2\times$ and $3\times$ cases are deliberately broadened sensitivity tests rather than independently calibrated alternative priors. The reported intervals remain conditional on the adopted \texttt{asfgrid}, $f_{\Delta\nu}$, and population-based $f_{\nu_{\max}}$ framework and do not exhaust all uncertainties in stellar input physics. The numerical stability of the resulting posterior summaries is demonstrated in Supplementary Fig.~S13. The scaling relations used in this framework can be written as \cite{kjeldsen1994amplitudes, chaplin2013asteroseismology, sharma2016stellar}:

\begin{equation}\label{equation:S1}
	\left(\frac{R}{R_{\odot}}\right)
	\simeq
	\left(\frac{\nu_{\max}}{f_{\nu_{\max}}\nu_{\max,\odot}}\right)
	\left(\frac{\Delta\nu}{f_{\Delta\nu}\Delta\nu_{\odot}}\right)^{-2}
	\left(\frac{T_{\rm eff}}{T_{{\rm eff},\odot}}\right)^{0.5},
\end{equation}
\begin{equation}\label{equation:S2}
	\left(\frac{M}{M_{\odot}}\right)
	\simeq
	\left(\frac{\nu_{\max}}{f_{\nu_{\max}}\nu_{\max,\odot}}\right)^{3}
	\left(\frac{\Delta\nu}{f_{\Delta\nu}\Delta\nu_{\odot}}\right)^{-4}
	\left(\frac{T_{\rm eff}}{T_{{\rm eff},\odot}}\right)^{1.5},
\end{equation}
\begin{equation}\label{equation:S3}
	\Delta\nu \simeq f_{\Delta\nu}\Delta\nu_{\odot}
	\left(\frac{\rho}{\rho_{\odot}}\right)^{0.5},
\end{equation}
\begin{equation}\label{equation:S4}
	\frac{\nu_{\max}}{\nu_{\max,\odot}}
	\simeq
	f_{\nu_{\max}}\frac{g}{g_{\odot}}
	\left(\frac{T_{\rm eff}}{T_{{\rm eff},\odot}}\right)^{-0.5}.
\end{equation}

The complete set of stellar parameters for each host is summarized in Table \ref{tab:para}; the corresponding baseline-prior \texttt{asfgrid} posterior distributions are shown in Supplementary Figs. S10--S12, and the prior-width validation is given in Supplementary Table S3. Their derived masses are combined with our radial-velocity measurements to infer planetary properties. We employed \texttt{PBjam 2.0} throughout the mode-identification and peakbagging analysis \cite{2025AJ....169..322N} to identify and peakbag the $\ell=0$, 1, and 2 oscillation patterns and to infer the dipole gravity-mode period spacing $\Delta\Pi_1$ and coupling factor $q$. For each target, we fitted the $\ell=0$ and $\ell=2$ modes, removed their model contributions from the working spectrum, and then fitted the remaining $\ell=1$ mixed-mode pattern. For the $\ell=1$ empirical-prior selection, only the observed $T_{\rm eff}$ and $\Delta\nu$ ranges were applied; the prior retains both RGB and CHeB period-spacing regimes. Because several dipole-model dimensions are absent from many library entries, we completed only those missing dimensions with explicitly stated broad priors; their distributions and numerical bounds are given in Supplementary Methods B. The two stages yield distinct posterior parameter sets, so the $\ell=0,2$ and $\ell=1$ corner plots are shown separately for each star. The seismic entries in Table~\ref{tab:para} use $\nu_{\max}$ and the central $\Delta\nu$ value from the $\ell=0,2$ analysis, while $\Delta\Pi_1$ and $q$ come from the $\ell=1$ analysis. We independently estimated the $\Delta\nu$ uncertainty by jointly fitting the peakbagged radial and quadrupole frequencies with a common curved acoustic pattern and propagating both the mode-frequency errors and the sensitivity to the finite fitted-order interval. The six joint posterior distributions are shown individually in Supplementary Figs. S3--S8; the dipole parameters and the independent $\Delta\nu$ check are listed in Supplementary Tables S1 and S2. Because the finite TESS time base and spectral window limit the security of the weakest individual peaks, the quoted parameter intervals are conditional on the retained \texttt{PBjam 2.0} mode pattern and the adopted two-cavity description rather than a marginalization over every possible discrete mode assignment; they do not include possible systematic error from unmodelled additional cavities or window-function effects. Nevertheless, the inferred $\Delta\Pi_1$ values and the positions of all three stars in the diagnostic diagrams (Fig. \ref{fig:deltapi-dnu}) place them securely in the red-clump regime.

\subsection*{Parameter Determination for the Three Exoplanets}

All three exoplanets in this study are confirmed detections via the radial-velocity method \cite{2018PASJ...70...59T, 2017A&A...602A..58J, 2015A&A...584A..79L}. The two derived planetary parameters---the minimum mass and the orbital semi-major axis---both depend on the host-star mass. Writing the edge-on minimum-mass solution as $m_{\min}=M_p\sin i$, we used the exact spectroscopic mass function and Kepler's third law \cite{mayor1995jupiter, lovis2010radial}:

\begin{equation}\label{equation:S7}
	\frac{P K^3}{2\pi G}\left(1-e^2\right)^{3/2}
	=\frac{m_{\min}^3}{\left(M_*+m_{\min}\right)^2},
\end{equation}

\begin{equation}\label{equation:S8}
	a = \left[\frac{G\left(M_*+m_{\min}\right)P^2}{4\pi^2}\right]^{1/3}.
\end{equation}

Here, $K$ is the radial-velocity (RV) semi-amplitude, $e$ the orbital eccentricity, and $P$ the orbital period. We refit all available RV data sets using the \texttt{exoplanet} package \cite{2021JOSS....6.3285F}, performing Markov chain Monte Carlo (MCMC) sampling to obtain posterior distributions for the fit parameters. For data sets containing multiple instruments or instrumental configurations, the model included independent velocity zero-point offsets and additional jitter terms for each data set. We then propagated the baseline stellar-mass posterior together with the asymmetric marginal RV constraints on $P$, $K$, and $e$ through Eqs.~\ref{equation:S7} and \ref{equation:S8} in $10^6$ Monte Carlo trials. Each stellar-mass posterior draw was used equally often, while the RV quantities were drawn from two-piece normal distributions constructed to reproduce their reported medians and 16th--84th percentile limits; draws outside the physical ranges $P>0$, $K>0$, and $0\leq e<1$ were rejected. In every trial we solved Eq.~\ref{equation:S7} numerically and included $m_{\min}$ in the total mass in Eq.~\ref{equation:S8}. The medians and 16th--84th percentiles of the resulting distributions are reported in Table~\ref{tab:para}; the other orbital entries are obtained directly from the RV fit. The phase-folded RV curves are shown in the main text (Fig. \ref{fig:DeltaPi}), whereas the posterior corner plots, RV curves and residuals, and fitted instrument-dependent parameters are provided in Supplementary Figs. S18--S23 and Supplementary Table S7.

The radial-velocity measurements for 24 Boo were obtained with the coudé-mounted High Dispersion Echelle Spectrograph (HIDES) on the 1.88 m reflector at Okayama Astrophysical Observatory (OAO), employing both the conventional slit mode (HIDES-S) and the fiber-fed modes before and after the upgrade (HIDES-F1 and HIDES-F2) \cite{2023PASJ...75.1030T}. The radial-velocity measurements for TYC 4282-605-1 were obtained with HARPS-N on the Telescopio Nazionale Galileo (TNG) and with GIANO at the TNG \cite{2017A&A...602A..58J}. The radial-velocity observations for HD 12648 were obtained with the high-resolution, fiber-fed Bohyunsan Observatory Echelle Spectrograph (BOES) mounted on the 1.8 m telescope at the Bohyunsan Optical Astronomy Observatory (BOAO) \cite{2015A&A...584A..79L}. All of the above radial-velocity datasets have been made publicly available in their respective published papers.

\subsection*{Stellar evolution models}

To test whether the planets could have survived the maximum expansion of their host stars during the preceding red-giant-branch evolution, we compared the measured planetary semi-major axes with tip-of-the-red-giant-branch (TRGB) radii. The primary comparison in Fig.~\ref{fig:HR} uses controlled MIST/MESA grids spanning $0.8\leq M_{\rm init}/M_\odot\leq1.8$ and $-1.0\leq[\mathrm{M/H}]\leq+0.3$. We calculated grids for $\alpha_{\rm MLT}=1.5$, 1.7, 1.82, 2.0, and 2.2, with $\alpha_{\rm MLT}=1.82$ as the solar-calibrated reference. For each mixing length, the two-dimensional radius grid was interpolated with a tensor-product PCHIP scheme. We then evaluated
\begin{equation}
R_{\rm TRGB}(M_{\rm init},[\mathrm{Fe/H}]_s;\alpha_{\rm MLT})
\quad\hbox{against}\quad a_{{\rm p},s}
\end{equation}
at the measured metallicity of each host $s$ for each of the five mixing-length parameters. Curves evaluated at the central metallicity are accompanied by envelopes spanning the quoted $[\mathrm{Fe/H}]\pm1\sigma$ interval. The measured orbit was converted from au to solar radii and plotted at the mapped initial mass, with both asymmetric uncertainties shown. Thus, $R_{\rm TRGB}>a_{\rm p}$ is read directly wherever a model curve lies above the orbital point.

We also repeated the fixed-metallicity comparison using six public, precomputed stellar-evolution libraries: nonrotating MIST v1.2 EEP tracks, PARSEC v1.2S scaled-solar tracks, updated BaSTI solar-scaled tracks, Dartmouth theoretical PreMS--RGB tracks with $[\alpha/{\rm Fe}]=0$ and the standard-helium relation, the public YREC EEP grid, and the 2016 YaPSI track family with $\alpha_{\rm MLT}=1.82126$ \cite{dotter2016mist,choi2016mist,bressan2012parsec,hidalgo2018basti,dotter2008dartmouth,demarque2008yrec,spada2017yapsi}. That comparison, together with representative MIST evolutionary tracks, is shown in Supplementary Fig. S15; the exact releases and adopted settings are listed in Supplementary Table S6. The public tracks were not modified, and their dispersion is interpreted as a combined inter-library systematic rather than a controlled change of one physical parameter. For 24 Boo the six-library comparison uses the conservative criterion of comparing $R_{\rm TRGB}$ with $2a_{\rm p}$; the physical $a_{\rm p}$ comparison is used for the other two hosts and for both controlled-physics families. The interpolation, grid-coverage audit, and probability calculation are described in Supplementary Methods D.

\subsection*{Planetary orbital evolution}

In a system consisting of a single star and a single planet, the onset of post–main-sequence evolution profoundly modifies the dynamical interaction between the two bodies. The orbital evolution is governed primarily by three processes:  
(i) mass loss from either the star or the planet,  
(ii) tidal dissipation induced by the stellar expansion, and  
(iii) hydrodynamic drag due to coupling with the stellar wind \cite{veras2016post}.  
The combined secular evolution of the planetary semi-major axis $a$ can be expressed as \cite{alexander1976orbital, livio1984star, villaver2009orbital}

\begin{equation}\label{equation:S9}
	\frac{{\rm d}a}{{\rm d}t} \;=\;
	-\,a\,\frac{\dot{M}_{\star} + \dot{M}_{p}}{M_{\star} + M_{p}}
	\;+\;\frac{2a}{M_{p}v}\,\bigl(F_{f} + F_{g}\bigr)
	\;+\;\left(\frac{{\rm d}a}{{\rm d}t}\right)_{\!{\rm tidal}} .
\end{equation}

Here $M_\star$ and $M_p$ denote the stellar and planetary masses, respectively, and overdots indicate time derivatives (so $\dot{M}<0$ represents mass loss).  
The term $v$ is the planet’s orbital velocity relative to the local stellar wind flow.  
The drag forces $F_f$ and $F_g$ correspond to the frictional and gravitational components of the wind–planet interaction, respectively \cite{villaver2009orbital}.

The first term represents the effect of changing the total system mass. Mass loss from either the star or the planet leads to an increase in the orbital semi-major axis $a$, reflecting the adiabatic expansion of the orbit as the gravitational potential weakens \cite{alexander1976orbital, livio1984star}. For evolved stars, substantial outflows are expected; empirical mass-loss prescriptions and wind models predict significant stellar mass-loss rates $\dot{M}_\star$ during the giant phases \cite{reimers1977absolute, lamers1999introduction}. Planetary mass may also decrease if the atmosphere is stripped by high-energy radiation or by the impacting stellar wind, contributing an additional term with $\dot{M}_p < 0$ \cite{rodriguez2019erosion}. However, in practice, the planetary mass loss is negligible compared to the stellar mass loss and can safely be ignored in our calculations.

The second term quantifies the aerodynamic and gravitational drag exerted by the stellar wind on the orbiting planet. This coupling extracts orbital energy and angular momentum, driving a gradual reduction in $a$. The magnitude of this effect depends sensitively on the local wind density and the relative velocity between the planetary motion and the wind flow \cite{villaver2009orbital}.

The final term, $\left({\rm d}a/{\rm d}t\right)_{\rm tidal}$, accounts for tidal dissipation. As the star evolves up the giant branches and its radius expands, the tidal torque typically intensifies and may dominate the late-stage orbital evolution of sufficiently close-in planets, resulting in orbital decay and, in extreme cases, planetary engulfment \cite{veras2016post}. This track is not used as a host-specific fit; rather, it provides a physically consistent example for identifying the generic critical-orbit behaviour expected for a planet that survives from the RGB to the RC phase.

The frictional (aerodynamic) drag acting on a planet moving through the stellar wind can be approximated as the drag on a solid sphere traveling through a dilute medium \cite{villaver2009orbital}:
\begin{equation}\label{equation:S10}
	F_{f} \;=\; -\,\frac{1}{2}\,C_{\rm D}\,\pi R_{\rm p}^{2}\,\rho_{\rm wind}\,v^{2},
\end{equation}
where $C_{\rm D}$ is the dimensionless drag coefficient, $R_{\rm p}$ is the planetary radius, $\rho_{\rm wind}$ denotes the local wind density, and $v$ is the relative velocity of the planet with respect to the stellar wind flow.

In addition to the direct aerodynamic drag, the planet’s gravity focuses the surrounding gas, producing a dynamical (gravitational) drag force on the body \cite{ostriker1999dynamical}.  
Following the notation adopted in planetary orbital-evolution calculations \cite{villaver2009orbital}, this effect can be expressed as
\begin{equation}\label{equation:S11}
	F_{\rm grav} \;=\; -\,\frac{4\pi\,(G M_p)^{2}\,\rho_{\rm wind}}{c_s^{2}}\,I(\mathcal{M}),
\end{equation}
where $c_{s}$ is the local sound speed of the wind, and $\mathcal{M} \equiv v / c_{s}$ is the Mach number.  
For the supersonic regime relevant here, we follow Villaver and Livio \cite{villaver2009orbital} and adopt $I(\mathcal{M}) \simeq 0.5$ in this notation.

For post–main-sequence giants, tides raised by a planetary companion are primarily dissipated through turbulent viscosity associated with large convective eddies in the stellar envelope \cite{zahn1977tidal, zahn1989tidal}.  
This mechanism converts orbital energy into heat, leading to a secular decay of the semi-major axis with a steep $(R_{\star}/a)^{8}$ dependence.  
The efficiency of this process is controlled by the convective eddy turnover timescale $\tau_{\rm conv}$ and the envelope mass $M_{\rm env}$ \cite{villaver2009orbital, mustill2012foretellings, veras2016post}:
\begin{equation}\label{equation:S12}
	\begin{aligned}
	\left(\frac{{\rm d}a}{{\rm d}t}\right)_{\rm tidal}
	&= -\frac{a}{9\,\tau_{\rm conv}}\,
	\frac{M_{\rm env}}{M_{\star}}\,
	\frac{M_{p}}{M_{\star}}\,
	\left(1+\frac{M_{p}}{M_{\star}}\right) \\
	&\quad \times
	\left(\frac{R_{\star}}{a}\right)^{8}
	\left[
	2p_{2}
	+ e^{2}\left(
	\frac{7}{8}p_{1}
	-10p_{2}
	+\frac{441}{8}p_{3}
	\right)
	\right],
	\end{aligned}
\end{equation}
\begin{equation}\label{equation:S13}
	\begin{aligned}
	\frac{{\rm d}e}{{\rm d}t}
	&= -\frac{a}{36\,\tau_{\rm conv}}\,
	\frac{M_{\rm env}}{M_{\star}}\,
	\frac{M_{p}}{M_{\star}}\,
	\left(1+\frac{M_{p}}{M_{\star}}\right) \\
	&\quad \times
	\left(\frac{R_{\star}}{a}\right)^{8}
	\left(
	\frac{5}{4}p_{1}
	-2p_{2}
	+\frac{147}{4}p_{3}
	\right),
	\end{aligned}
\end{equation}
\begin{equation}\label{equation:S14}
	\tau_{\rm conv} \;=\;
	\left[\frac{M_{\rm env}\,\bigl(R_{\star}-R_{\rm env}\bigr)}{3\,L_{\star}}\right]^{1/3},
\end{equation}
\begin{equation}\label{equation:S15}
	p_{l}\;\approx\;\frac{9}{2}\,
	\min\!\left\{\,1,\;\frac{4\pi^{2}a^{3}}{l^{2}\,G\,(M_{\star}+M_{p})\,\tau_{\rm conv}^{2}}\,\right\},
	\qquad l=1,2,3.
\end{equation}

\eqref{equation:S9}–\eqref{equation:S15} together describe the long-term evolution of the semi-major axis $a$ under three coupled effects:  
(i) mass loss,  
(ii) hydrodynamic drag from the stellar wind, and  
(iii) tidal dissipation.  
Once the planet’s initial semi-major axis and physical parameters are specified, and a self-consistent stellar-evolution model for the host star is provided, the orbital history can be fully integrated forward in time.

To complete the system of equations, we introduce the following physically motivated assumptions.  
The planet in the illustrative calculation is treated as a gas giant with fixed mass and radius.  
Although atmospheric escape could, in principle, decrease $M_{p}$ over time, the expected mass loss is negligible compared to the host star’s mass loss during the giant phases; thus, we set $\dot{M}_{p}=0$ in \eqref{equation:S9}.  
In \eqref{equation:S9}–\eqref{equation:S10}, $v$ denotes the planet’s velocity relative to the stellar wind.  
Assuming the wind outflow is orthogonal to the planet’s instantaneous orbital motion, we take
\[
v \;=\; \sqrt{\,v_{\rm orb}^{2}+v_{\rm wind}^{2}\,}\,.
\]
For the frictional drag term, we adopt $C_{\rm D}=0.9$, appropriate for a sphere moving through a uniform, rarefied medium.  
The wind density at the planet’s orbital distance is given by \cite{villaver2009orbital}
\begin{equation}\label{equation:S16}
	\rho_{\rm wind} \;=\; \frac{\dot{M}_{\star}}{4\pi\,a^{2}\,v_{\rm wind}}\,,
\end{equation}
where we adopt a characteristic red-giant wind velocity of $v_{\rm wind}=5~\mathrm{km\,s^{-1}}$ \cite{villaver2009orbital}.  
For the gravitational-drag component, the local sound speed is evaluated using the ideal-gas relation
\begin{equation}\label{equation:S17}
	c_{s} \;=\; \sqrt{\frac{\gamma\,k_{\rm B}\,T}{\mu\,m_{\rm p}}}\,,
\end{equation}
where $\gamma$ is the adiabatic index (we take $\gamma=5/3$ for a monatomic gas), $k_{\rm B}$ is Boltzmann’s constant, $T$ is the wind temperature, $\mu$ is the mean molecular weight, and $m_{\rm p}$ is the proton mass.  
Observations indicate that red-giant winds have temperatures in the range $1650$–$3800$~K, varying with radial distance \cite{o2020alma}; for simplicity, we adopt a constant value of $T=2500$~K.  
Numerical simulations further show that, for orbits outside the stellar radius, frictional drag generally dominates over gravitational drag; only very close to the photosphere does gravitational drag become comparable, at which point tidal dissipation typically drives the orbital decay.  
Hence, the adoption of an isothermal wind approximation introduces negligible uncertainty into our results.
The remaining stellar quantities required for the illustrative orbital-evolution calculation were extracted from the representative evolutionary model. The stellar mass-loss rate, $\dot{M}_\star(t)$, was computed as the time derivative of the stellar mass along the track, while the instantaneous stellar mass, radius, and luminosity were obtained from the model output. The convective-envelope properties required for the tidal prescription, including the envelope mass $M_{\rm env}$ and the radius at the base of the envelope $R_{\rm env}$, were taken from the corresponding internal structure profiles.

We integrated a sequence of test-planet orbits from the main sequence through the RGB, TRGB, and RC phases of this representative model. For each trial initial semi-major axis, the integration was classified according to whether the planet was engulfed before the TRGB or survived into the RC phase. We then used a binary-search refinement to converge on the critical initial semi-major axis that just survives the TRGB expansion. The resulting critical orbit approaches the maximum stellar radius at the TRGB, demonstrating that, in the canonical single-star orbital-evolution framework, a planet that survives into the RC phase should not end up on an orbit smaller than the host star's TRGB radius.

}

\showmatmethods{} % Display the Materials and Methods section

\dataavail{TESS light curves of 3 host stars processed by the SPOC pipeline are available from MAST (\url{https://archive.stsci.edu/}). 
		We used \texttt{exoplanet} (\url{https://docs.exoplanet.codes/en/latest/}) to fit the stellar radial-velocity curves. The MESA stellar evolution models are available at \url{https://docs.mesastar.org/en/latest/}. The public precomputed tracks used for the six-library TRGB-radius comparison can be accessed directly from MIST (\url{https://mist.science/model_grids.html}), PARSEC (\url{https://stev.oapd.inaf.it/PARSEC/tracks_v12s.html}), BaSTI (\url{http://basti-iac.oa-abruzzo.inaf.it/tracks.html}), Dartmouth (\url{https://rcweb.dartmouth.edu/stellar/grid_old.html}), YREC (\url{https://doi.org/10.5281/zenodo.14908017}), and YaPSI (\url{https://www.astro.yale.edu/yapsi/download_grids.html}). Seismic masses and radii were corrected using \texttt{asfgrid} (\url{https://www.physics.usyd.edu.au/k2gap/Asfgrid/}); the $\nu_{\max}$ and central $\Delta\nu$ values were obtained from the \texttt{PBjam 2.0} radial--quadrupole analysis, with the $\Delta\nu$ uncertainty independently evaluated from the peakbagged mode frequencies; and oscillation modes were identified and peakbagged throughout with \texttt{PBjam 2.0} (\url{https://doi.org/10.3847/1538-3881/adcb37}). The representative MESA inlist and the complete model files for every node of the controlled MIST/MESA grids, together with the associated analysis outputs and radial-velocity fitting products, are deposited in Figshare at \url{https://doi.org/10.6084/m9.figshare.33691738}; these files fully reproduce the Supplementary Methods D tests of $\alpha_{\rm MLT}$ and $\Delta Y/\Delta Z$. }

\acknow{We gratefully acknowledge Dr. Teng Huan-Yu for providing the radial-velocity measurements of 24 Boo.
\textbf{Funding:} This work is partly supported by the National Natural Science Foundation of China (Nos. 12603101, grant recipient: LWX; 12573038, grant recipient: LWP; 12303040, grant recipient: LFX; and 12503040, grant recipient: SQB.); the Postdoctoral Fellowship Program of CPSF (Nos. GZB20260766, grant recipient: LWX; and GZC20252095, grant recipient: SQB); the Science Foundation of Yunnan Province (Nos. 202501AS070055, grant recipient: QSB; 202401AS070046 and 202503AP140013, grant recipient: ZLY; and 202401AT070143, grant recipient: LFX); the 2022 CAS ``Light of West China'' Program (grant recipient: LWP); the Yunnan Revitalization Talent Support Program; the China Manned Space Program (Grant No. CMS-CSST-2025-A16); the International Partnership Program of the Chinese Academy of Sciences (No. 020GJHZ2023030GC); and the China Postdoctoral Science Foundation (No. 2025M773194, grant recipient: SQB).}

\showacknow{} % Display the acknowledgments section

\clearpage
\newpage
\bibsplit[33]
%Use \bibsplit to split the references from the body of the text. Value "[3]" represents the number of reference in the left column (Note: Please avoid single column figures & tables on this page.)

% Bibliography
\bibliography{pnas-sample}

\end{document}